\documentclass[conference]{IEEEtran}
\IEEEoverridecommandlockouts

\usepackage{amsmath}
\usepackage{subcaption}
\usepackage{graphicx}
\usepackage{comment}
\usepackage{tipa}
\usepackage{xurl}
\usepackage{bm}
\usepackage{booktabs}
\usepackage{xcolor} 

\title{An Interpretable Acoustic Coordinate System for Japanese Consonant Morae in Dysarthric Speech}

\author{
\IEEEauthorblockN{
Yuma Enomoto\IEEEauthorrefmark{2}, 
Misa Yamaguchi\IEEEauthorrefmark{3}, 
Hiroki Nishikawa\IEEEauthorrefmark{2}, 
Ittetsu Taniguchi\IEEEauthorrefmark{2}, \\
Kazuhide Matsunaga\IEEEauthorrefmark{3}, 
Narikazu Uzawa\IEEEauthorrefmark{3}, 
Takao Onoye\IEEEauthorrefmark{2}
}
\IEEEauthorblockA{\IEEEauthorrefmark{2}Graduate School of Information Science and Technology, The University of Osaka, Japan}
\IEEEauthorblockA{\IEEEauthorrefmark{3}Graduate School of Dentistry, The University of Osaka, Japan}
\IEEEauthorblockA{Email: \{enomoto.yuma, nishikawa.hiroki, i-tanigu, onoye\}@ist.osaka-u.ac.jp, \\\{u558580a@ecs., matsunaga.kazuhide.dent@, uzawa.narikazu.dent@\}osaka-u.ac.jp}
}

\begin{document}
\maketitle

\begin{abstract}
Dysarthria reduces speech intelligibility, requiring self-directed rehabilitation based on fine-grained feedback to capture gradual articulatory changes. For Japanese speakers, such feedback is naturally provided in mora units due to its distinct moraic timing and consonant–vowel structure. While recent visualization methods have explored distance-based feedback, most are limited to vowels. Extending them to consonants is challenging, as generative embeddings often smooth transient spectral structures. We propose a visualization system for Japanese consonant morae that maps articulatory deviations onto a two-dimensional space defined by temporal and spectral distances from healthy reference speech. Experiments show consistent geometric separation of consonant categories, with Silhouette Coefficients exceeding 0.5 and origin-based ROC-AUC scores above 0.86. Furthermore, evaluations using real dysarthric speech achieved a high ROC-AUC of 0.897 in screening for unclear pronunciations. Perturbation tests confirm orthogonal temporal and spectral shifts, yielding a physically interpretable mapping for rehabilitation.
\end{abstract}


\section{Introduction}\label{introduction}
Dysarthria is a motor speech disorder that degrades speech intelligibility and reduces patients’ quality of life. Structural deficits after surgical resection for oral cancer, for example, can impair articulatory function and reduce communicative ability \cite{bressmann2004consonant, lirani2015dysarthria}. Because rehabilitation often continues beyond clinical sessions, patients need feedback that can quantify articulatory changes and present them in an interpretable form \cite{ribeiro19_interspeech}. Speech technology can support this process by providing acoustic feedback for self-directed training \cite{benway2024artificial}. In current clinical practice, however, rehabilitation is commonly monitored through therapists’ perceptual judgments \cite{de1997test}, which are susceptible to inter-rater variability and provide limited support for tracking gradual changes in articulation \cite{kent1996hearing}. Toward self-directed speech rehabilitation, this study focuses on an interpretable assessment of how a patient’s speech differs from healthy reference speech. We construct an acoustic representation that visualizes mora-level deviations as a basis for rehabilitation feedback.

Existing objective feedback methods do not adequately describe how dysarthric speech deviates from a target articulation. Text transcription can indicate whether an utterance is recognized correctly, but it does not specify the acoustic direction or magnitude of the deviation. Distance-based visualization has been studied as an alternative feedback approach because it can present deviations from reference speech in an acoustic feature space \cite{lee2016relationship, kuo2023relationship}. Michizoe et al. \cite{michizoe2024japanese} showed that vowel visualization can present the acoustic distance between a patient’s speech and a target reference. However, this approach has not been extended to consonants, which pose a different signal-processing problem. Consonants include short-duration and nonstationary acoustic cues, such as closures, bursts \cite{shadle1991effect}, frication \cite{pruthi2004acoustic}, and rapid transitions \cite{o1957acoustic}. These cues can be smoothed in generic embedding spaces, and low-dimensional projections may lose the physical meaning needed for rehabilitation feedback \cite{wattenberg2016use}.

This paper addresses the problem of constructing an interpretable acoustic representation for Japanese consonant morae. Japanese speech is organized in mora units, and each consonant mora contains both timing-related and spectral cues \cite{sugai2017mental}. We use this structure to decompose consonant-mora deviations into temporal and spectral distances from healthy reference speech. The resulting two-dimensional space provides an origin-referenced representation. The distance from the origin represents the degree of acoustic deviation from healthy reference speech, while the two axes are designed to separately reflect timing-related and spectral-structure-related deviations.

The contributions of this study are fourfold. First, we formulate a mora-based temporal--spectral visualization framework for Japanese consonant speech. Second, we show that the proposed space separates consonant categories in healthy speech more clearly than generative embedding baselines while preserving physically defined axes. Third, we use simulated dysarthric speech to verify that rhythmic and spectral perturbations affect different directions in the proposed space. Fourth, we evaluate real dysarthric speech and show that the origin-referenced distance is associated with unclear pronunciations, suggesting its potential as an objective assessment cue for rehabilitation-oriented feedback.

\section{Related Work}\label{related work}
\subsection{Category-based Speech Assessment}\label{relate-auto}
Recent advancements in automated assessment have focused on classifying dysarthric speech into pre-defined categories. One primary approach is error detection, which identifies specific articulatory errors 
\cite{shen2022self, cao24b_interspeech}. Another major approach involves classifying the severity of dysarthria, which has successfully categorized patients into specific severity levels
\cite{riosurrego23_interspeech, javanmardi2023wav2vec,zaheera24_interspeech,kadirvelu2025speaker}.

While category-based assessment methods are effective for clinical diagnosis \cite{darley1969differential}, self-directed rehabilitation introduces additional requirements. Although accurate for their intended diagnostic purposes, categorical labels are not designed to capture the granular articulatory improvements crucial for sustained patient motivation \cite{maas2008principles}. Because a class-based output remains unchanged unless a patient's improvement crosses a specific severity boundary, patients may struggle to recognize slight acoustic progress. A continuous, distance-based representation enables patients to observe gradual movement toward the target even when categorical labels remain unchanged. Such a representation also provides an intuitive quantitative reference that patients can use for self-monitoring \cite{byun2012investigating, preston2014ultrasound}, which traditional classification systems are not designed to offer.

\subsection{Distance-based Speech Visualization}\label{relate-vowel}
To address the limitation where category-based assessment fails to capture granular articulatory improvements, recent work has explored the distance-based visualization of speech features in a latent space \cite{sainburg2019latent, tits19_interspeech}. When targeting Japanese speech, it is important to consider its specific phonological structure. The basic unit of Japanese speech is the mora, which generally consists of a consonant followed by a vowel. Since Japanese speakers perceive and produce speech based on these moraic units \cite{otake1993mora, cutler1994mora}, developing a mora-based visualization method is important for intuitive speech assessment. Aligned with this perspective, Michizoe et al. \cite{michizoe2024japanese} proposed a method using a Variational Autoencoder (VAE) \cite{kingma2013auto} to project Japanese vowel morae onto a 2D plane, allowing patients to intuitively grasp the positional relationship between their own speech and the target speech clusters. This hypothesis is empirically examined in our separability experiment in Section \ref{subsection:4-2}.

However, considering the acoustic characteristics of consonants, challenges exist in extending this method to the visualization of consonant morae. Consonants are characterized by complex spectral fine structures \cite{shadle1991effect, pruthi2004acoustic, o1957acoustic}. 
VAE-based methods typically employ reconstruction objectives with regularization terms, which may reduce the discriminability of transient consonantal features in low-dimensional projections by smoothing fine-grained spectral structures \cite{blaauw16_interspeech}. Consequently, when performing dimensionality reduction for visualization on a 2D plane, the model prioritizes reconstructing the more stable global structure of vowels, which generally have larger energy and longer duration within the mora \cite{beckman1982segment, han1994acoustic}. This tendency can result in the attenuation of fine-grained features that are important for consonant identification \cite{kumar2019melgan}.

\section{Interpretable Two-Dimensional Temporal–Spectral Visualization}\label{proposed_method}

\subsection{System Overview}
\begin{figure*}[t]
  \centering
  \includegraphics[width=0.85\linewidth]{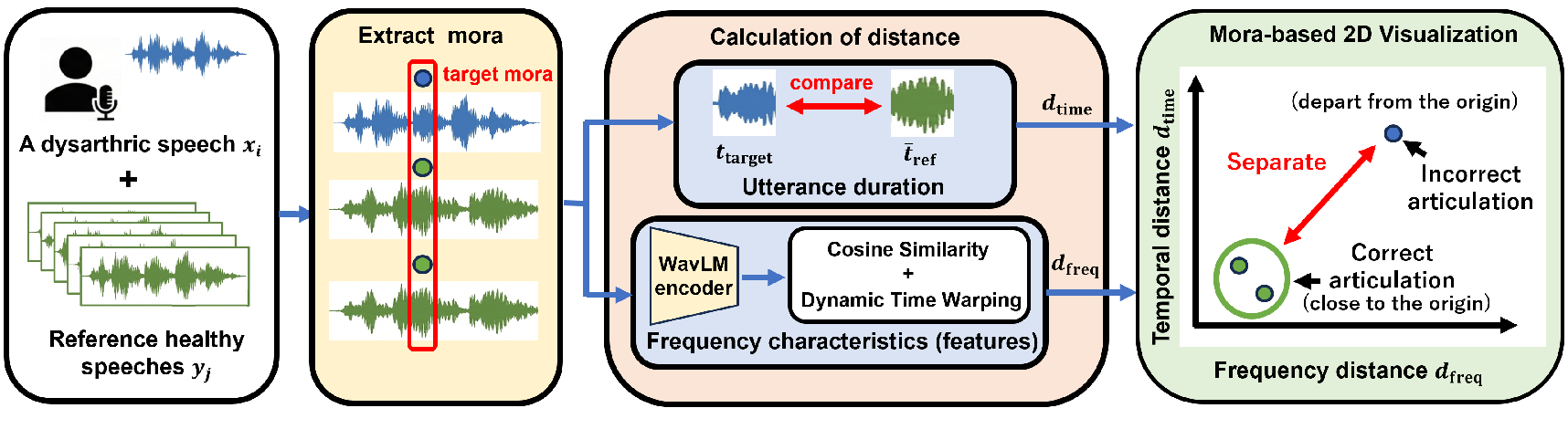}
  \caption{Overview of the proposed system. Temporal distances $d_{\mathrm{time}}$ and frequency  distances $d_{\mathrm{freq}}$ between patient ($\bm{x}_i$) and healthy ($\bm{y}_j$) speech are calculated to visualize articulatory deviations on a two-dimensional plane.}
  \label{fig:system_overview}
\end{figure*}

As illustrated in Figure \ref{fig:system_overview}, the proposed framework consists of four main stages: input processing, target mora extraction, distance calculation, and mora-based 2D visualization.

In the input stage, the system accepts a patient's speech and a reference group of healthy speech signals. Specifically, let $\bm{x}_i$ $(i=1, \dots, I)$ be the discrete speech signal of the $i$-th patient, and $\bm{y}_j$ $(j=1, \dots, J)$ be the discrete speech signal of the $j$-th reference healthy speaker. We assume these signals are accompanied by pre-computed mora-level alignments, denoting the start and end timestamps for the $k$-th mora in the patient's speech as $(s_{\bm{x}_i, k}, e_{\bm{x}_i, k})$.

In the target mora extraction stage, the system identifies the specific mora to be evaluated. Instead of physically cropping the raw audio waveforms, this step isolates the temporal position of the target mora based on the timestamps. This ensures that the acoustic context is fully preserved when processing the utterance in the subsequent stage \cite{choi2026self}.

Next, the system computes two distinct distance metrics to evaluate the articulatory deviation. First, it calculates a Temporal Distance $d_{\mathrm{time}}$ by comparing the utterance duration between $\bm{x}_i$ and $\bm{y}_j$. Second, it computes a Frequency Distance $d_{\mathrm{freq}}$ based on frequency characteristics. This is achieved by extracting high-dimensional representations via a WavLM encoder \cite{chen2022wavlm} and applying Dynamic Time Warping with cosine similarity.

Finally, in the “Mora-based 2D Visualization” stage, the two calculated scalar distances are used as coordinates to project the target mora onto a 2D plane. Because the origin represents zero deviation from the healthy reference mean, correctly articulated morae are expected to appear closer to the origin, although natural inter-speaker variability may result in some dispersion. In contrast, articulations with larger deviations are projected to depart from the origin.
This spatial separation can help patients recognize whether the deviation is primarily temporal or spectral, rather than just the overall magnitude.

\subsection{Temporal Distance for Rhythmic Deviations}
The temporal distance quantifies the rhythmic accuracy of the utterance. In dysarthric speech, reduced motor function often leads to extreme prolongation of specific, difficult-to-articulate morae, or irregular duration fluctuations, as seen in ataxic dysarthria \cite{darley1969differential}. To capture these unnatural duration variations localized in specific morae, a simple absolute difference in duration is insufficient because overall speech speed varies even among healthy speakers. Thus, we utilize a duration ratio to consistently evaluate the degree of rhythmic distortion regardless of the individual's baseline speaking rate.

We define the temporal distance $d_{\mathrm{time}, i, k}$ for the $k$-th mora of patient $i$ as the absolute natural logarithm of the ratio between the patient's mora duration and the average duration of the healthy speakers:
\begin{equation}
    d_{\mathrm{time}, i, k} = \left| \ln \left( \frac{t_{\mathrm{target}, i, k}}{\overline{t}_{\mathrm{ref}, k}} \right) \right|,
\end{equation}
where $t_{\mathrm{target}, i, k} = e_{\bm{x}_i, k} - s_{\bm{x}_i, k}$ is the duration of the patient's mora, and $\overline{t}_{\mathrm{ref}, k}$ is the average duration of the $k$-th mora across all healthy speakers $j=1 \dots J$. This logarithmic formulation ensures that symmetrical deviations are evaluated as equivalent rhythmic errors.

\subsection{Spectral Distance for Structural Deviations}
The frequency distance quantifies the degree of acoustic deviation caused by atypical physical articulation, measured in a high-dimensional spectral representation space.

Physical constraints in articulators distort the vocal tract's resonance and the resulting speech spectrum. To capture these structural deviations, we utilize the intermediate layers of WavLM \cite{chen2022wavlm}. Recent findings demonstrate that specific dimensions within WavLM embeddings strongly correlate with physical acoustic cues tied to vocal tract resonances, such as formant characteristics \cite{sivalingam2025dimension}. While early model layers retain irrelevant non-articulatory noise and higher layers overly abstract speech into discrete phonetic units, the intermediate transitional space preserves these physical structural properties \cite{pasad2023comparative}. Extracting these intermediate representations allows us to evaluate atypical articulatory dynamics.

Let $\bm{X}_i = f_\text{WavLM}(\bm{x}_i)$ and $\bm{Y}_j = f_\text{WavLM}(\bm{y}_j)$ denote the full feature sequences extracted from the $i$-th patient and the $j$-th healthy speaker, respectively. Using the mora alignment $(s_{\bm{x}_i, k}, e_{\bm{x}_i, k})$, we extract the corresponding feature subsequence for the $k$-th mora. We denote the patient's subsequence as $\hat{\bm{X}}_{i,k} = (\bm{u}_1, \dots, \bm{u}_{T})$ and the healthy speaker's subsequence as $\hat{\bm{Y}}_{j,k} = (\bm{v}_1, \dots, \bm{v}_{T'})$, where $\bm{u}$ and $\bm{v}$ are frame-level feature vectors. This extraction captures high-level acoustic-linguistic representations, preserving dynamic phonetic variations over time.

To evaluate the acoustic deviation while accounting for temporal fluctuations such as consonant duration differences, we calculate the distance using Dynamic Time Warping (DTW) \cite{sakoe2003dynamic}. The distance between the patient's sequence $\hat{\bm{X}}_{i,k}$ and a reference sequence $\hat{\bm{Y}}_{j,k}$ is defined as the minimum accumulated cost along the optimal alignment path $\mathcal{W}$:
\begin{equation}
    f_{\mathrm{freq}}(\hat{\bm{X}}_{i,k}, \hat{\bm{Y}}_{j,k}) = \min_{\mathcal{W}} \left( \frac{1}{|\mathcal{W}|} \sum_{(t, t') \in \mathcal{W}} d_{\cos}(\bm{u}_t, \bm{v}_{t'}) \right).
\end{equation}

We employ the cosine distance as the local cost function: $d_{\cos}(\bm{u}_t, \bm{v}_{t'}) = 1 - \frac{\bm{u}_t \cdot \bm{v}_{t'}}{\lVert\bm{u}_t\rVert \lVert \bm{v}_{t'}\rVert}$. Since cosine similarity depends only on the vector angle, this metric ensures robustness against variations in voice loudness. Combined with DTW, which absorbs temporal differences, this frequency distance isolates and emphasizes articulatory deviations that are independent of timing errors, although minor alignment artifacts may remain \cite{dehak2010cosine}.

Finally, the frequency distance $d_{\mathrm{freq}, i, k}$ is computed by averaging the distances to all healthy speakers $j=1 \dots J$ to eliminate bias from individual voice characteristics:
\begin{equation}
    d_{\mathrm{freq}, i, k} = \frac{1}{J} \sum_{j=1}^{J} f_{\mathrm{freq}}(\hat{\bm{X}}_{i,k}, \hat{\bm{Y}}_{j,k}).
\end{equation}

\section{Experiments}
\subsection{Setup}\label{Setup}
To evaluate the proposed visualization system, we conducted three distinct experiments: (1) an assessment of the foundational consonant separability using healthy speech, (2) a verification of the geometric interpretability of the 2D plane using simulated dysarthric speech, and (3) an evaluation to quantify articulatory degradation and discriminate intelligibility using clinical dysarthric speech.

We utilized a Japanese speech dataset comprising 17 healthy speakers (6 males, 11 females; ages 28--66) and 5 patients with dysarthria. The study protocol was approved by the Ethics Committee of the University of Osaka (Protocol ID: R7-E16). All five patients are male, with individual ages of 25, 27, 47, 49, and 64 years. Their shared primary etiology is subtotal resection of the tongue. The healthy speakers read a passage from ``The North Wind and the Sun" three times in Japanese, whereas each patient provided a single recording of the same passage. Environmental noise and respiration sounds were reduced using AudioDirector 365 \cite{AudioDirector365}. To prevent forced alignment degradation, the audio was split into three short sentences \cite{moreno1998recursive}. We utilized the Montreal Forced Aligner (MFA) v3.3.8 \cite{mcauliffe17_interspeech} to obtain phone-level timestamps, which were subsequently merged into mora units. For the healthy speakers, the first two repetitions served as the Reference Audio Set, while their third repetition was used for evaluation. Conversely, the single recording from each patient was used exclusively for evaluation. All audio was downsampled to 16 kHz.

We extracted features from the 6th Transformer layer of the WavLM Base Plus model \cite{chen2022wavlm}, utilizing it as an intermediate representation that effectively captures structural articulatory properties. To mitigate speaker-dependent bias and exclude unclear articulations, we selected 20 reference samples per mora from the Reference Audio Set that minimized the cumulative feature distance, keeping the evaluation data unseen. We evaluated the separation performance for 14 consonant morae: /ka, so, ta, na, hi, ma, yo, ra, wa, ga, ze, da, ba, pi/. We define the target consonant morae as $C_{\mathrm{same}}$ and morae with different consonants as $C_{\mathrm{others}}$. We computed the mean Silhouette Coefficient (SC) for $C_{\mathrm{same}}$ against $C_{\mathrm{others}}$ to assess cluster separability. Additionally, to assess origin-based discriminability, we computed ROC-AUC scores using the Euclidean distance from the origin as an indicator of articulatory deviations. The 95\% confidence intervals (CIs) for both metrics were estimated via bootstrapping with 1,000 resamples.

To verify the effectiveness of the proposed DTW-based distance metrics, we implemented two baselines. The first is a CNN-based VAE trained on $ 64 \times 64$ Mel-spectrograms to map morae into a 2D latent space.The second is t-SNE, applied directly to the 6th-layer WavLM features as a representative non-linear dimensionality reduction method. 


Furthermore, for the explainability experiment, we simulated specific factors of speech unintelligibility commonly observed in dysarthria to verify the geometric independence of the temporal and spectral axes. While actual dysarthric speech often involves compounded errors, this controlled simulation aims to isolate individual articulatory factors to observe their independent effects on the visualized plane. We prepared two augmented conditions using the mora /ka/ based on distinct signal processing techniques.

For the temporal perturbation denoted as $C_{\mathrm{temp\_change}}$, we employed Time-Scale Modification (TSM) \cite{vachhani2018data} to stretch the time domain while preserving the original frequency components. This technique maps the original time $t$ to the modified time $t' = \frac{1}{\alpha} \cdot t$, where $\alpha$ is the tempo factor. By setting $\alpha=0.6$, we effectively reproduced the reduced speech rate commonly caused by weakened articulatory muscles without unintentionally shifting the pitch. 

For the spectral perturbation denoted as $C_{\mathrm{freq\_change}}$, we applied Frequency Warping \cite{osipov2025dysarthria} to the Short-Time Fourier Transform (STFT) spectrogram of the speech signal. The modified spectrogram amplitude $S'(f, t)$ is generated by displacing the frequency bins according to $S'(f, t) = S(f + \Delta f(f), t)$. Specifically, we applied a uniform sine-wave-based displacement across the data, defined as $\Delta f(f) = \epsilon \cdot \sin(2\pi N f + \theta)$. This uniform sinusoidal modulation shifts the speech formants and blurs the spectral components, mimicking the unclear pronunciation of consonants and vowels caused by structural tissue loss, such as partial glossectomy. We utilized a displacement magnitude of $\epsilon=0.020$ (with fixed parameters $N$ and $\theta$) to represent medium-severity structural distortions without altering the duration.

Finally, for the clinical intelligibility experiment, we defined ``unclear" pronunciation at the mora level using an Automatic Speech Recognition (ASR) approach. The patient speech samples were transcribed using the UDTalk \cite{UDTalk} ASR engine; a mora was labeled as ``Clear" if the recognized phonemes matched the target transcript and ``Unclear" otherwise. To evaluate the system's performance, we analyzed the distance from the origin using two methods. First, we computed the unnormalized mean Euclidean distance across individual visualization planes to assess the scale of articulatory deviation for each mora. Second, to evaluate origin-based discriminability across the dataset, we integrated all mora data into a unified plane using Z-score normalization based on healthy reference statistics. Using this normalized plane, we computed the ROC-AUC score to quantify how well the distance from the origin screens for speech degradation.


\subsection{Assessment of Consonant Separability}\label{subsection:4-2}

\begin{table}[t]
    \caption{Clustering performance (Silhouette Coefficient of $C_{\mathrm{same}}$) with 95\% confidence intervals.}
    \label{tab:integrated_results}
    \centering
    \resizebox{\columnwidth}{!}{
    \begin{tabular}{lccc}
        \toprule
        Mora & Baseline (VAE) & Baseline (t-SNE) & Proposed \\
        \midrule
        /ka/ & 0.24 [0.21, 0.27] & 0.78 [0.77, 0.79] & 0.62 [0.59, 0.65] \\
        /so/ & 0.38 [0.24, 0.50] & 0.94 [0.94, 0.95] & 0.85 [0.84, 0.87] \\
        /ta/ & 0.02 [-0.01, 0.05] & 0.20 [0.20, 0.21] & 0.07 [0.00, 0.13] \\
        /na/ & 0.33 [0.21, 0.43] & 0.96 [0.95, 0.96] & 0.68 [0.60, 0.73] \\
        /hi/ & 0.64 [0.51, 0.73] & 0.94 [0.94, 0.95] & 0.81 [0.77, 0.84] \\
        /ma/ & 0.38 [0.36, 0.40] & 0.88 [0.87, 0.88] & 0.66 [0.63, 0.68] \\
        /yo/ & 0.51 [0.48, 0.53] & 0.93 [0.92, 0.93] & 0.82 [0.79, 0.84] \\
        /ra/ & 0.53 [0.48, 0.58] & 0.85 [0.84, 0.85] & 0.64 [0.59, 0.69] \\
        /wa/ & 0.25 [0.16, 0.34] & 0.92 [0.91, 0.93] & 0.51 [0.32, 0.65] \\
        /ga/ & 0.14 [0.09, 0.19] & 0.69 [0.69, 0.70] & 0.21 [0.16, 0.25] \\
        /ze/ & 0.15 [0.11, 0.18] & 0.81 [0.80, 0.82] & 0.73 [0.70, 0.75] \\
        /da/ & 0.54 [0.48, 0.58] & 0.95 [0.95, 0.96] & 0.83 [0.81, 0.85] \\
        /ba/ & 0.20 [0.16, 0.23] & 0.91 [0.87, 0.93] & 0.69 [0.61, 0.75] \\
        /pi/ & 0.27 [0.11, 0.40] & 0.93 [0.92, 0.93] & 0.67 [0.55, 0.75] \\
        \bottomrule
    \end{tabular}
    }
\end{table}

\begin{figure*}[t]
    \centering
    \begin{subfigure}[b]{0.25\linewidth}
        \centering
        \includegraphics[width=\linewidth]{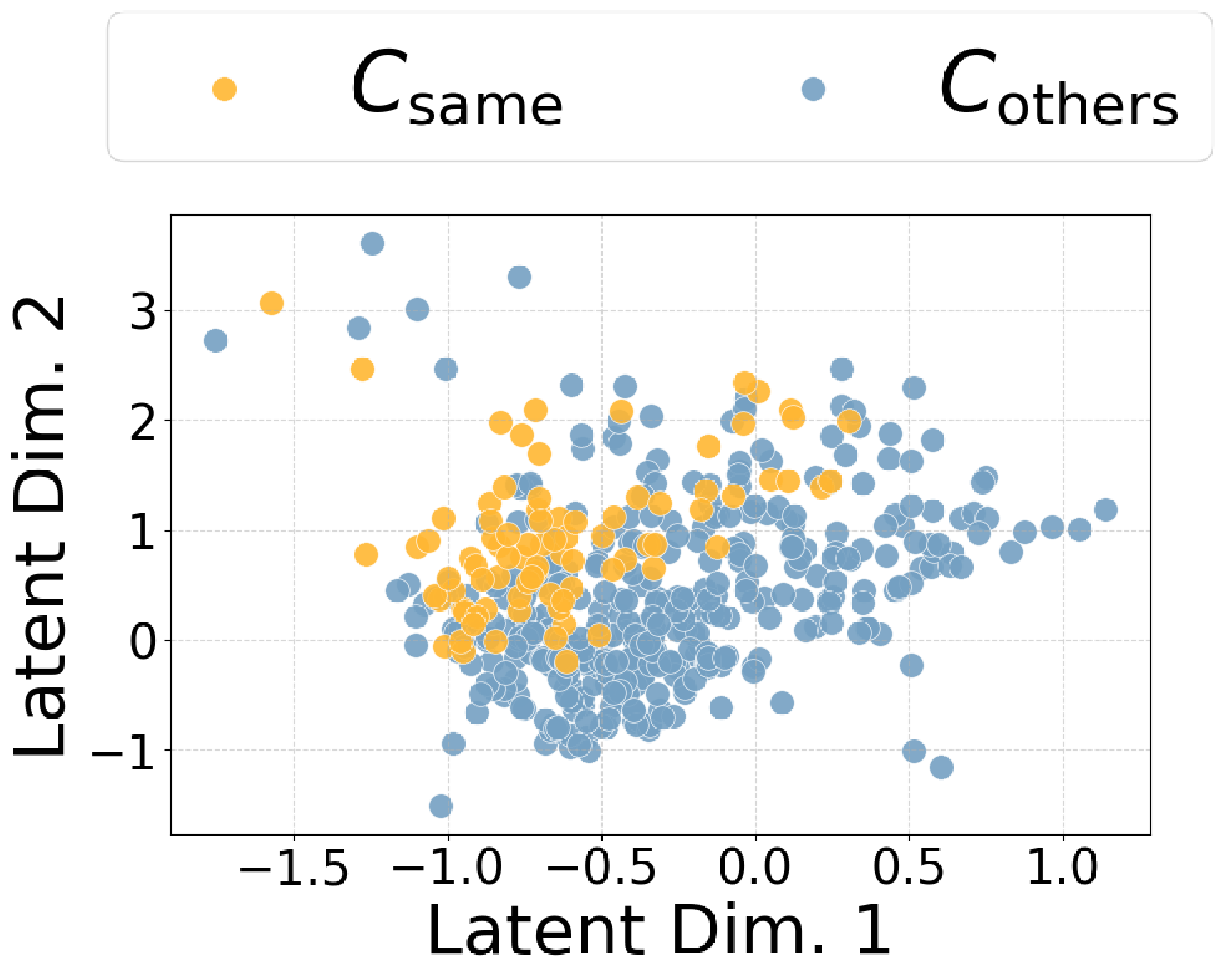}
        \caption{VAE}
        \label{sfig:vae_ka}
    \end{subfigure}
    \hfill
    \begin{subfigure}[b]{0.25\linewidth}
        \centering
        \includegraphics[width=\linewidth]{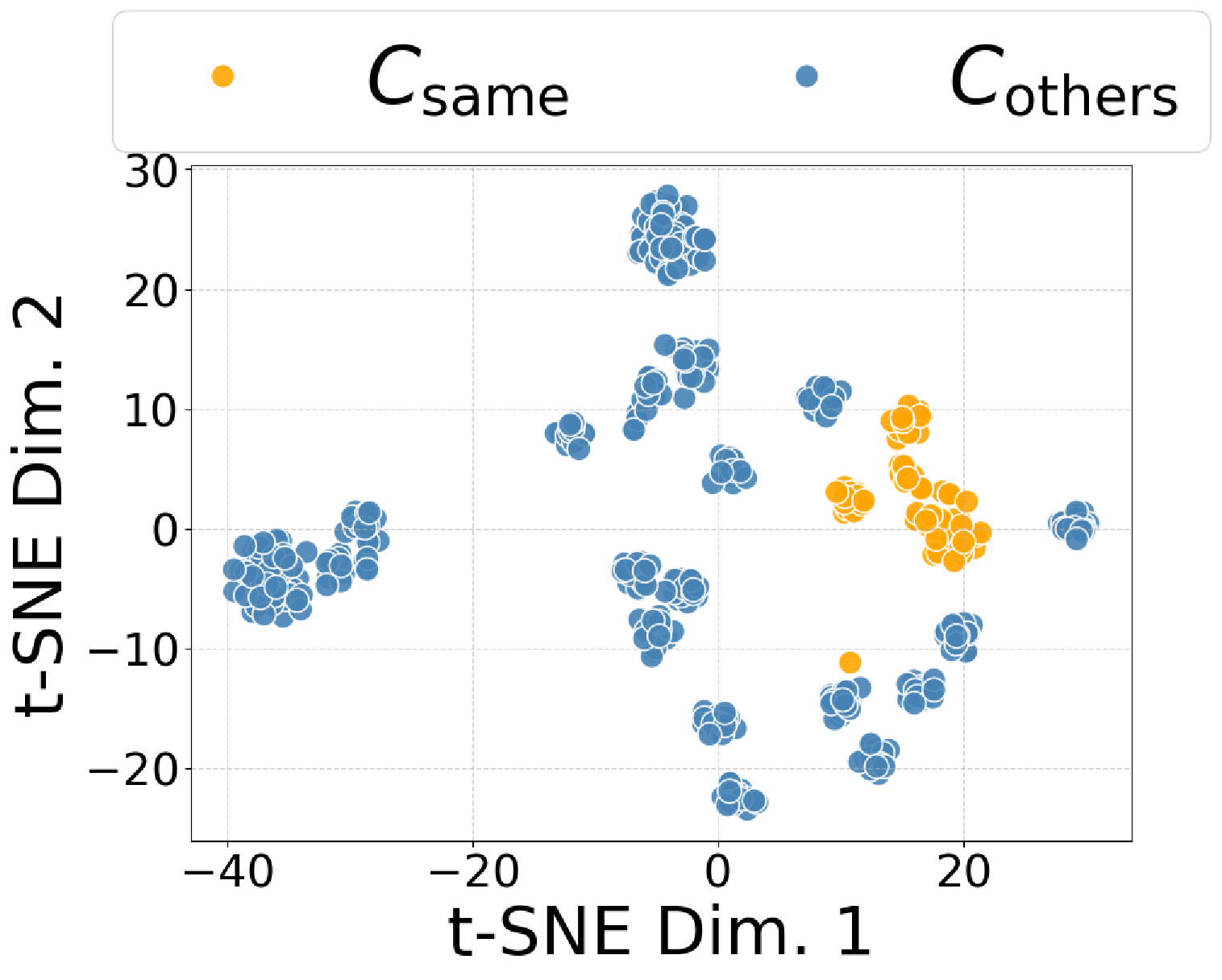}
        \caption{t-SNE}
        \label{sfig:tsne_ka}
    \end{subfigure}
    \hfill
    \begin{subfigure}[b]{0.25\linewidth}
        \centering
        \includegraphics[width=\linewidth]{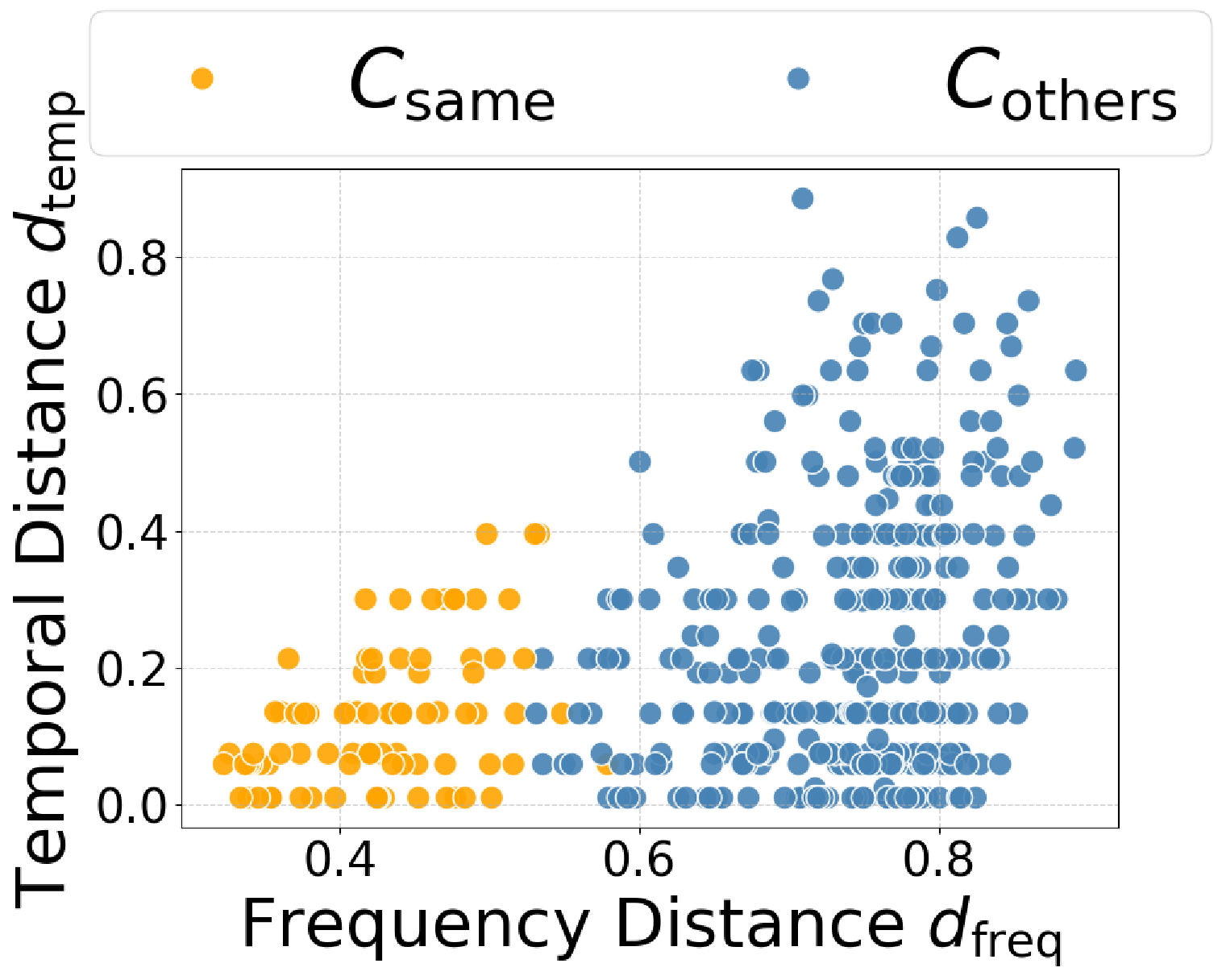}
        \caption{Proposed}
        \label{sfig:proposed_ka}
    \end{subfigure}
    \caption{Visualization of /ka/. (a) VAE overlaps significantly due to over-smoothing. (b) The t-SNE baseline separates the target well, but lacks physically meaningful axes. (c) Proposed method separates the target class $C_{\mathrm{same}}$ from the different consonant class $C_{\mathrm{others}}$.}
    \label{fig:comparison_ka}
\end{figure*}

Table \ref{tab:integrated_results} and Figure \ref{fig:comparison_ka} evaluate the separation of $C_{\mathrm{same}}$ against $C_{\mathrm{others}}$. As seen in Fig.~\ref{fig:comparison_ka}(a), the VAE baseline fails to separate the classes, yielding low Silhouette Coefficients (SC) and corroborating that generative models over-smooth fine consonantal features. In contrast, t-SNE and the proposed method mitigate this overlap, achieving SCs exceeding 0.5 and 95\% CIs consistently higher than those of the VAE, suggesting a robust improvement over the generative embedding in this setting.

Crucially, unlike t-SNE's physically meaningless latent space (Fig.~\ref{fig:comparison_ka}(b)), our proposed method projects features onto a 2D plane defined by explicit temporal and frequency distances (Fig.~\ref{fig:comparison_ka}(c)). In this space, the origin $(0,0)$ serves as the acoustic reference point for the target mora. The ideal representation is achieved when target utterances are clustered tightly near the origin, indicating high acoustic consistency. In contrast, utterances with different consonants are pushed to distant regions. As shown in Fig.~\ref{fig:comparison_ka}(c), our method successfully achieves this ideal distribution. This result demonstrates that our system preserves transient consonantal features via WavLM, projecting these differences onto explicit temporal and frequency axes.

Regarding origin-based discriminability, Table \ref{tab:roc_auc_proposed} shows that the proposed method achieved high ROC-AUC scores across all 14 evaluated morae. Even for challenging cases with high intra-phoneme variance, the scores exceeded 0.86, while many other morae reached a perfect score of 1.00. These results support the idea that the distance from the origin in our 2D space can serve as a quantitative indicator of articulatory deviation.

\begin{table}[t]
    \caption{Mean ROC-AUC scores and their 95\% CIs using the proposed method.}
    \label{tab:roc_auc_proposed}
    \centering
    \resizebox{\columnwidth}{!}{
    \begin{tabular}{lc|lc}
        \toprule
        Mora & ROC-AUC [95\% CI] & Mora & ROC-AUC [95\% CI] \\
        \midrule
        /ka/ & 1.00 [0.99, 1.00] & /ra/ & 0.98 [0.95, 1.00] \\
        /so/ & 1.00 [1.00, 1.00] & /wa/ & 0.96 [0.88, 1.00] \\
        /ta/ & 0.86 [0.82, 0.90] & /ga/ & 0.87 [0.79, 0.93] \\
        /na/ & 1.00 [1.00, 1.00] & /ze/ & 1.00 [1.00, 1.00] \\
        /hi/ & 1.00 [1.00, 1.00] & /da/ & 1.00 [1.00, 1.00] \\
        /ma/ & 1.00 [1.00, 1.00] & /ba/ & 0.99 [0.98, 1.00] \\
        /yo/ & 1.00 [1.00, 1.00] & /pi/ & 1.00 [0.98, 1.00] \\
        \bottomrule
    \end{tabular}
    }
\end{table}

\begin{figure}[t]
    \centering
    \begin{minipage}[b]{0.49\columnwidth}
        \centering
        \centerline{\includegraphics[width=\linewidth]{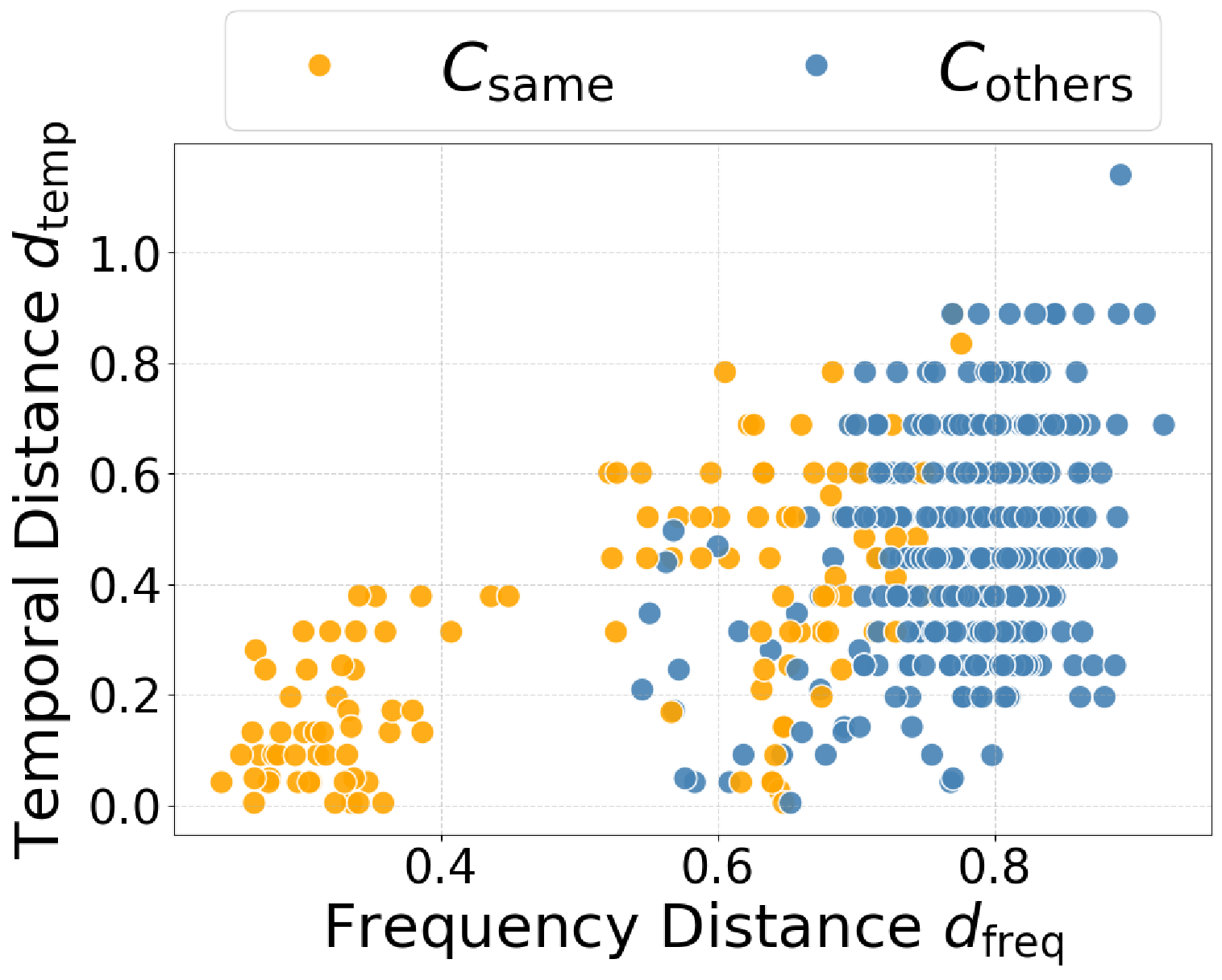}}
        \vspace{1mm}
        \centerline{(a) /ta/}
    \end{minipage}
    \hfill
    \begin{minipage}[b]{0.49\columnwidth}
        \centering
        \centerline{\includegraphics[width=\linewidth]{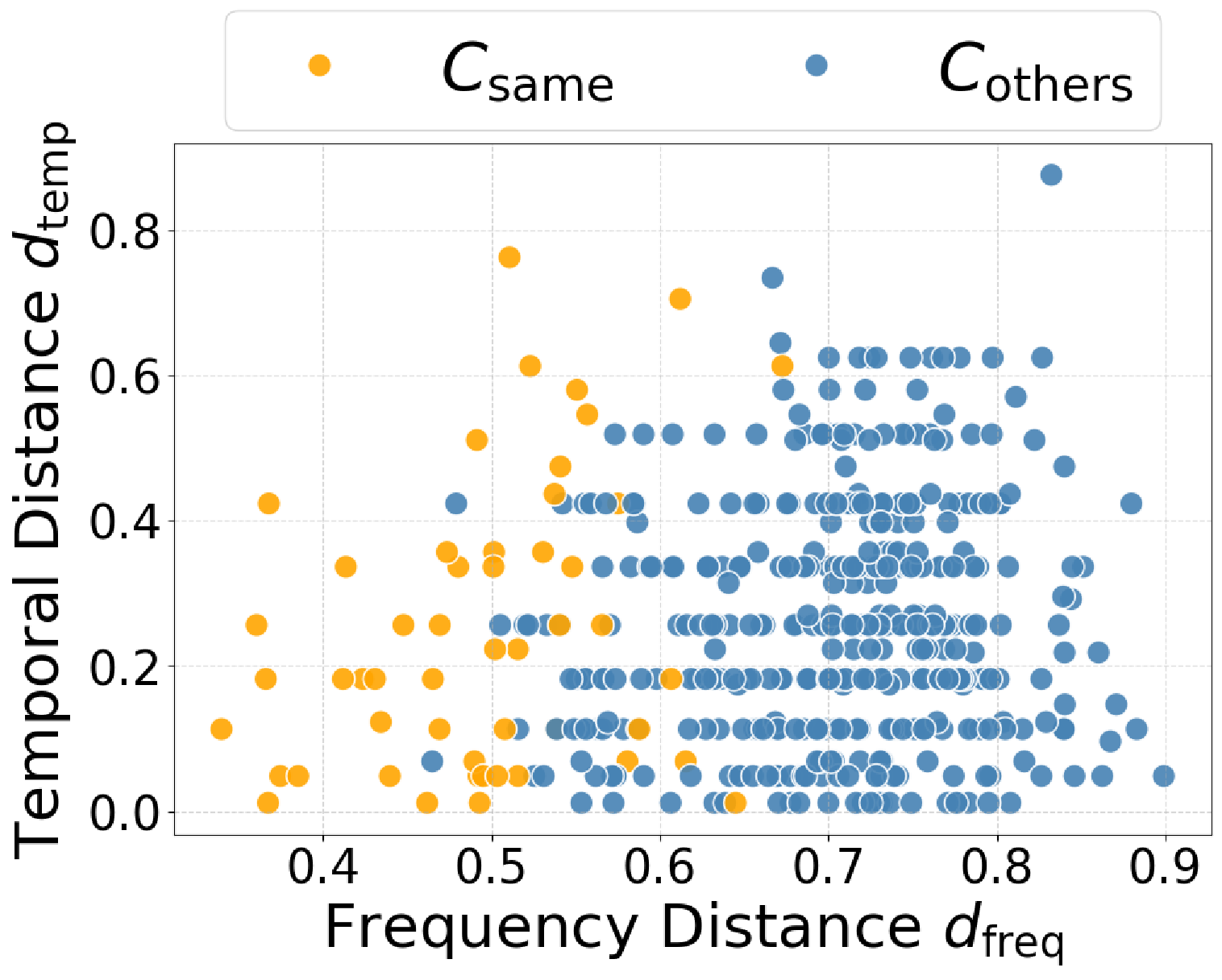}}
        \vspace{1mm}
        \centerline{(b) /ga/}
    \end{minipage}
    \caption{Visualization of /ta/ and /ga/. The distribution of /ta/ is bifurcated due to acoustic differences between sentence-final and word-medial occurrences (a). The variance of /ga/ is large due to its varying roles as a particle or part of a noun (b).}
    \label{fig:ta_ga_dist}
\end{figure}

Conversely, specific morae like /ta/ and /ga/ exhibited lower SCs, highlighting the system's sensitivity to linguistic contexts. As shown in Fig.~\ref{fig:ta_ga_dist}, the target class $C_{\mathrm{same}}$ fails to form cohesive clusters; /ta/ bifurcates (a) and /ga/ scatters (b). This fragmentation occurs because the target class mixes distinct linguistic contexts, such as varying word positions or allophonic variations. Consequently, target samples that differ in context from the dominant reference samples receive large distance penalties, causing overlap with other morae. These results suggest the importance of explicitly accounting for the linguistic context to accommodate acceptable intra-phoneme variations.

\subsection{Geometric Interpretability}

\begin{figure}[t]
    \centering
    \begin{subfigure}[b]{0.49\columnwidth}
        \centering
        \includegraphics[width=\linewidth]{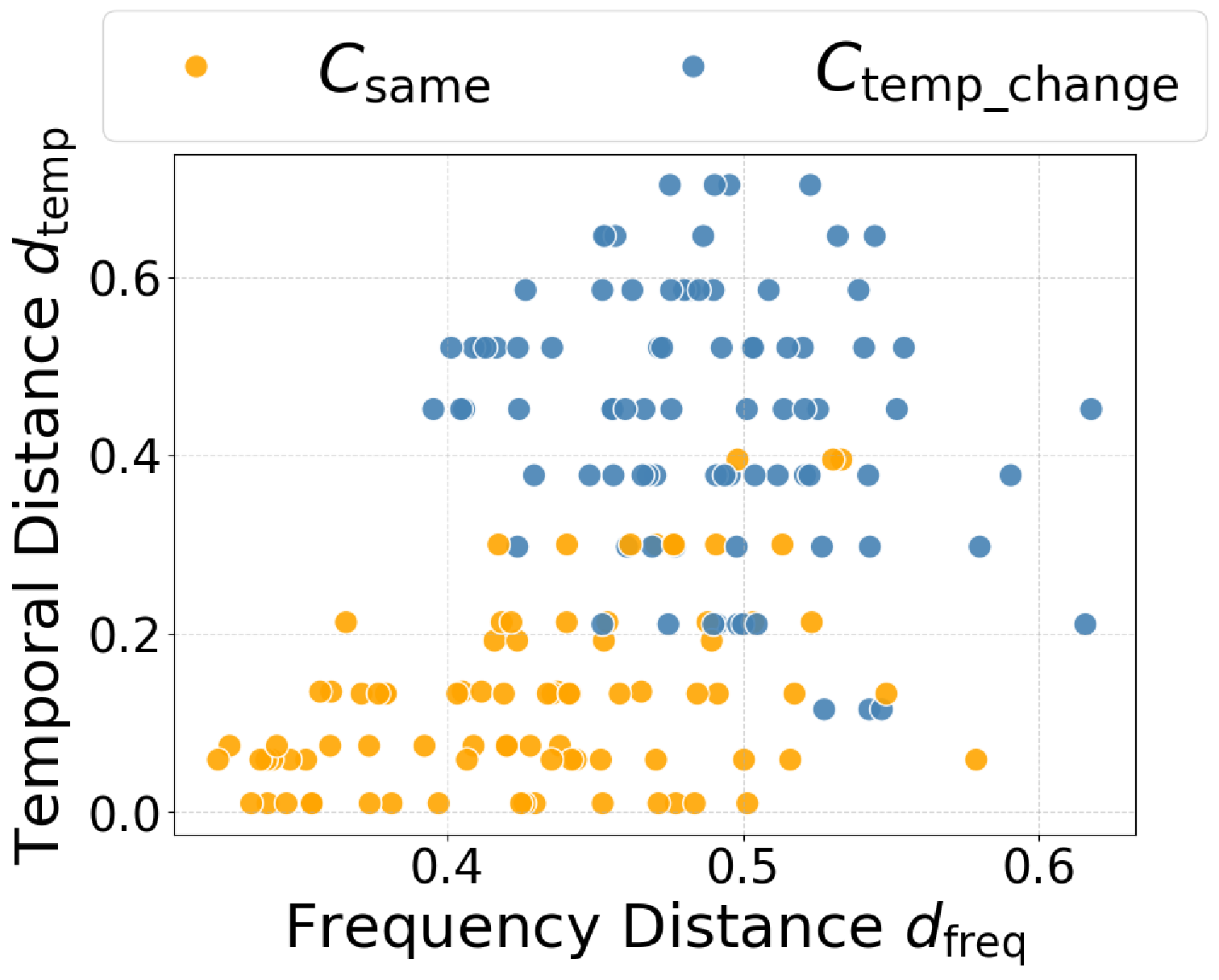}
        \caption{Temporal}
        \label{fig:pert_temp}
    \end{subfigure}
    \hfill
    \begin{subfigure}[b]{0.49\columnwidth}
        \centering
        \includegraphics[width=\linewidth]{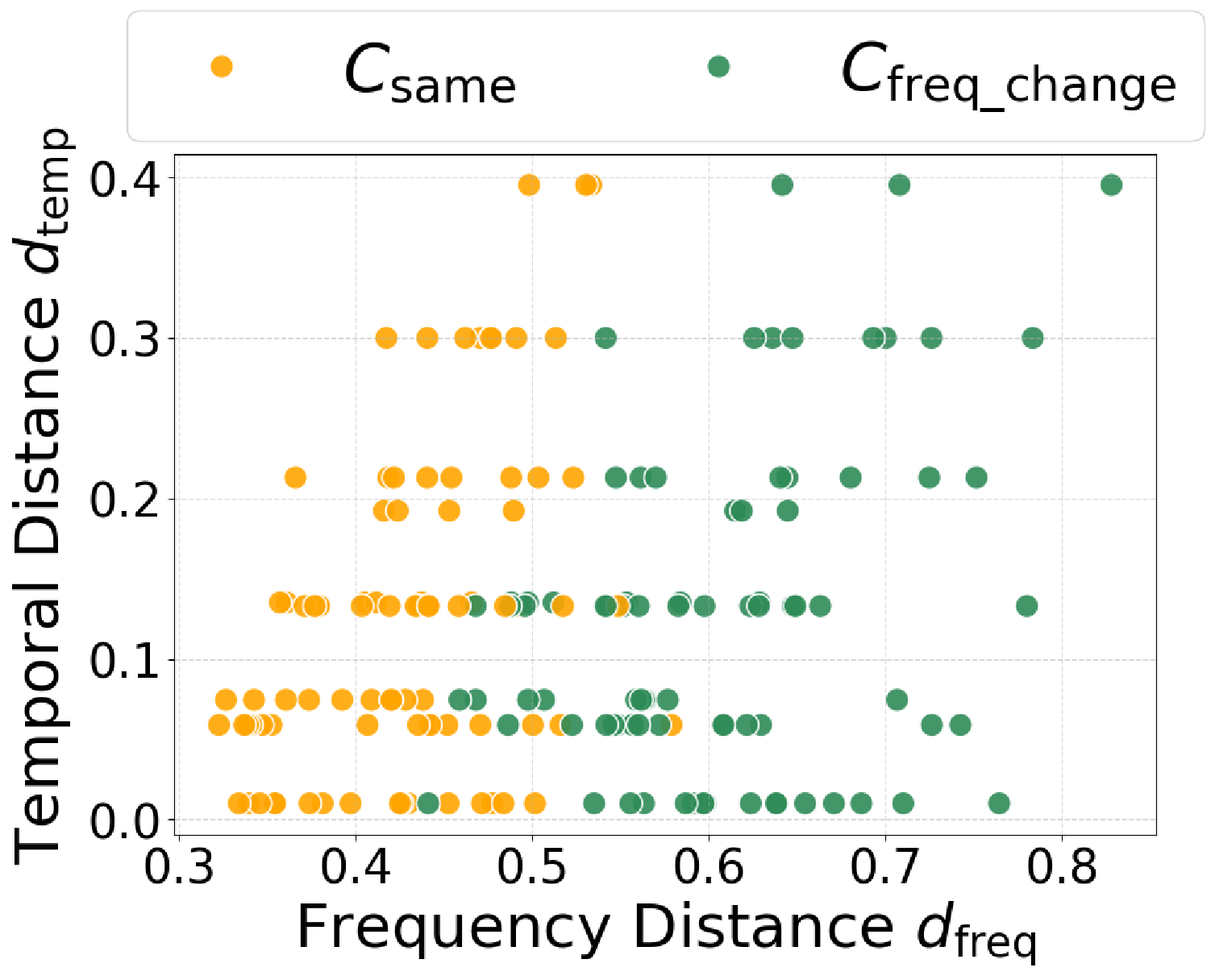}
        \caption{Frequency}
        \label{fig:pert_freq}
    \end{subfigure}
    \caption{Visualization of simulated dysarthric speech for /ka/. (a) Time-stretched samples $C_{\mathrm{temp\_change}}$ shift vertically, reflecting temporal deviations. (b) Frequency warping samples $C_{\mathrm{freq\_change}}$ shift horizontally, reflecting spectral structural deviations.}
    \label{fig:perturbation}
\end{figure}

To verify whether the 2D plane provides interpretable feedback for articulatory errors, Figure \ref{fig:perturbation} illustrates the base healthy samples ($C_{\mathrm{same}}$) alongside the simulated dysarthric speech defined in Section \ref{Setup}. The plots demonstrate that time stretching and frequency warping independently affect their respective axes.

As shown in Fig.~\ref{fig:perturbation}(a), relative to the centroid of the base samples $C_{\mathrm{same}}$, the time-stretched samples $C_{\mathrm{temp\_change}}$ shifted almost exclusively along the vertical axis (+0.32). The horizontal shift (+0.06), attributed to minor DTW alignment artifacts, is sufficiently small to confirm practical independence. Conversely, Fig.~\ref{fig:perturbation}(b) demonstrates that  the frequency-warped samples ($C_{\mathrm{freq\_change}}$) shifted exclusively along the horizontal axis (+0.18) with no temporal movement. Applying both perturbations yielded a diagonal shift (+0.32, +0.18) from $C_{\mathrm{same}}$, precisely reflecting the combined orthogonal displacements.

These independent movements support the geometric interpretability of the proposed 2D plane. Because the vertical and horizontal axes are designed to reflect rhythm-related and spectral structural changes, respectively, each deviation can be interpreted as indicating a dominant direction of acoustic difference. This geometric association suggests the system's potential to provide intuitive, feedback cues for self-directed rehabilitation. Crucially, this orthogonal separation may help users understand, addressing the limitations of conventional opaque error scores. By calculating the frequency distance using DTW and cosine similarity, the system intentionally absorbs variations in voice volume and timing, thereby emphasizing structural differences in the acoustic representation. For example, if a patient's pronunciation shifts exclusively along the horizontal axis, the feedback suggests that their rhythmic timing is correct, and they should focus their practice solely on adjusting physical articulator postures, such as tongue placement. By presenting speech deviations as geometrically separated dimensions, the framework empowers patients to understand not only that an error occurred, but which aspect of their speech should be examined further.

\subsection{Intelligibility Evaluation using Dysarthric Speech}

To evaluate whether the proposed system can capture the degree of articulatory degradation, we first analyzed the unnormalized Euclidean distance from the origin for both ``Clear" and ``Unclear" pronunciations. An independent t-test showed a significant difference between the groups ($p < 0.01$). The mean distance for the Clear group was 0.407, whereas the Unclear group exhibited a larger mean distance of 0.709. These results indicate that unclear articulations tend to be projected farther from the healthy reference origin.

\begin{figure}[t]
    \centering
    \includegraphics[width=0.75\linewidth]{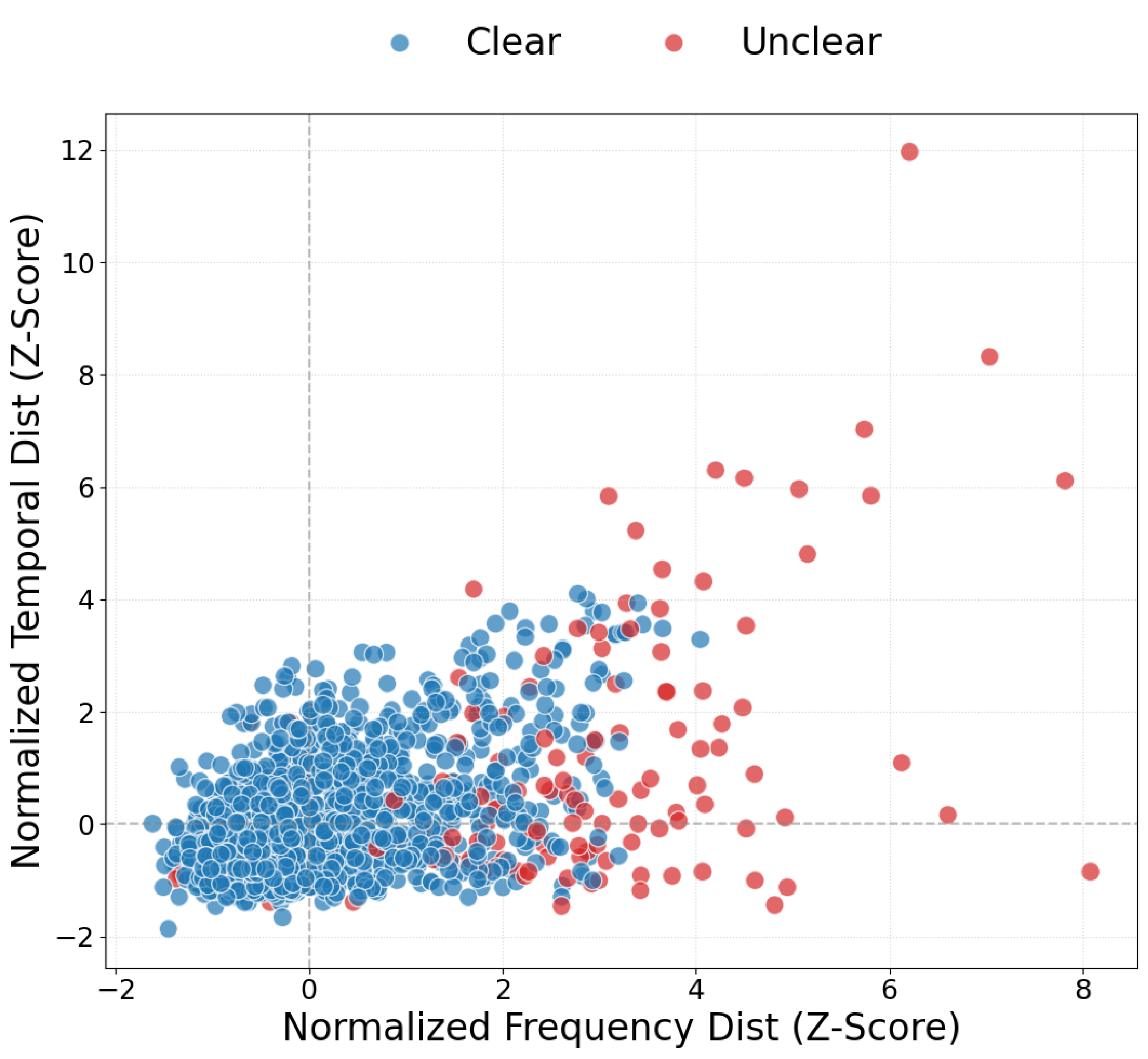} 
    \caption{Pooled 2D visualization of all evaluated morae across both healthy speakers and patients after Z-score normalization. The ``Clear" pronunciations (blue) tightly cluster near the origin $(0,0)$, whereas the ``Unclear" pronunciations (red) disperse outward. This distribution demonstrates that the geometric distance from the origin effectively captures the degree of articulatory degradation.}
    \label{fig:pooled_scatter}
\end{figure}

Furthermore, we assessed the system's practical screening capability across all consonant morae using the Z-score normalized unified plane. Figure \ref{fig:pooled_scatter} illustrates the distribution of the integrated data, showing a clear tendency for Unclear pronunciations to disperse outward from the central Clear cluster. Based on this normalized spatial distribution, we evaluated the origin-based discriminability by calculating the ROC-AUC score. The system achieved a high AUC of 0.897. This result suggests that the normalized geometric distance from the origin can indicate speech deviations.

These findings demonstrate the feasibility of the proposed two-dimensional visualization framework. Conventional ASR-based assessments often provide binary feedback, which fails to capture subtle acoustic improvements. In contrast, our results show that the continuous distance from the origin can serve as a quantitative cue for estimating the severity of the misarticulation. By mapping unclear pronunciations as geometric departures from the healthy reference, the system may help patients perceive their acoustic deviations more intuitively. This capability to visualize the degree of error could support patient motivation during self-directed rehabilitation, as it enables the tracking of gradual acoustic progress even before a categorical improvement is achieved.

\section{Conclusions}
In this paper, we proposed a 2D visualization system for Japanese consonant morae as an interpretable assessment basis for future self-directed speech rehabilitation. The system represents mora-level deviations from healthy reference speech using temporal and frequency distances, aiming to separately reflect timing-related and spectral-structure-related differences. Evaluations using Silhouette Coefficients and ROC-AUC scores showed that the proposed space provides origin-based discriminability for consonant morae. Perturbation experiments further suggested that the temporal and frequency dimensions respond in different directions under controlled modifications, supporting the intended design of the coordinate system. Furthermore, Evaluation using actual dysarthric speech showed that the geometric distance from the origin was associated with unclear pronunciations and achieved high discriminative performance. Context-dependent morae also revealed sensitivity to acceptable intra-phoneme variations. Future work will implement a context-aware reference selection mechanism and validate whether the proposed visualization can support rehabilitation feedback across a larger cohort of patients encompassing a wider variety of surgical cases.

\section*{Acknowledgements}
This work was partly supported by JSPS KAKEN Grant Number 25K12970.

\bibliographystyle{IEEEtran}
\bibliography{mybib}

\end{document}